\documentclass[a4paper,twocolumn,english,aps,prl]{revtex4}
\usepackage[T1]{fontenc}
\usepackage[latin1]{inputenc}
\usepackage{float}
\usepackage{graphicx,color}

\makeatletter

\providecommand{\LyX}{L\kern-.1667em\lower.25em\hbox{Y}\kern-.125emX\@}

\usepackage{babel}
\makeatother
\begin{document}

\preprint{This line only printed with preprint option}

\title{Probing ultra-cold Fermi atoms with a single ion}

\author{Y. Sherkunov}

\affiliation{Department of Physics, University of Warwick, Coventry, CV4 7AL,
UK}

\author{B. Muzykantskii}

\affiliation{Department of Physics, University of Warwick, Coventry, CV4 7AL,
UK}

\author{N. d'Ambrumenil}

\affiliation{Department of Physics, University of Warwick, Coventry, CV4 7AL,
UK}

\author{B.D. Simons}

\affiliation{Cavendish Laboratory, University of Cambridge, 
Cambridge, CB3 0HE, UK}



\begin{abstract}
We show that the recently proposed ionic microscope set-up [Kollath et.al, PRA, \textbf{76}, 063602 (2007)] could be 
adapted to measure the \textit {local single-particle energy distribution} of a degenerate Fermi gas 
\textit{in situ} with the resolution on the nanometer scale. 
We study an ion held in a Paul trap in an atomic Fermi gas
and compute the two-photon
Raman photo-association rate of the ion and an atom. We show that, as a function
of the detunings between the frequencies 
of the two incident lasers and 
 energies in the atom-ion system,
the photo-association rate directly measures the single-particle energy distribution in the Fermi
gas around the ion. We describe an experiment to measure the photo-association
rate of a trapped ion and argue that, 
as the position of the ion can be scanned through the Fermi gas,
this experiment directly  probes  the local energy and
spin-state distribution of the Fermi gas.

\end{abstract}
\maketitle

\section{introduction}

Most experimental techniques used to study 
ultra-cold  atomic Fermi systems yield either 
information about the momentum or single-particle energy distribution, 
averaged over a large part of a sample, or local information about the
number density.  
For example,
time-of-flight measurements of  a gas
after release from a trap have been used
to measure the column momentum distribution within the gas
\cite{Regal05,Chen06,Tarruell07}, while \emph{in situ} imaging methods 
have been used to study spatial variations in the density. 
It has been possible, for example, to demonstrate how a 
trapped Fermi gas with imbalanced spin populations separates 
into a fully paired superfluid, 
partially polarized normal liquid and fully polarized normal liquid 
as a function of increasing radius \cite{Shin06,Partridge06}. 
However, the spatial resolution of the \emph{in situ} optical imaging is limited by the wavelength
of the light and it has not yet been possible to determine whether a narrow region
of  FFLO phase (in which pairs have 
non-zero center of mass momentum \cite{Fulde64, Larkin65})
existed in the Fermi gas between the superfluid and partially 
polarized normal liquid. Enhanced spatial resolution has been obtained in
some systems by allowing the gas to expand ballistically before imaging.
This is how details of  the vortex structure 
in rotating systems were observed \cite{Zwierein2005, Zwierein2006}. 

One of the most powerful methods --- radio-frequency (RF) spectroscopy --- is used to probe single-particle energy distribution in Fermi gases. This method, introduced to study interaction effects in ultracold Fermi gases \cite{RegalJin03,Gupta03}, has been successfully used to observe the pairing gap in strongly interacting Fermi systems \cite{Chin04}. The momentum-resolved RF spectroscopy has been recently used to probe the single-particle spectral function $A(\mathbf{k},E)$ of a fermion with momentum $\mathbf{k}$ and energy $E$ \cite{Stewart08}. For review of modern progress in RF spectroscopy technique see \cite{Chen08}. The spatial resolution in RF spectroscopy is obviously limited by the  wavelength of light which does not allow to make local probes with spatial resolution better than $1\mu m$ \cite{Shin07}.

Another important tool for studying ultra-cold gases experimentally uses two-photon
Raman spectroscopy. It has been used successfully to induce optical Feshbach
resonances and probe ultra-cold gases in optical latices
\cite{Wynar00, Rom04, Theis04, Thalhammer05, Ryu05,Prodan03}.
Recently, Ko\u{s}trun and C\^ot\'e  suggested using two-photon Raman spectroscopy
to measure the temperature of a non-interacting Fermi gas \cite{KC03}
and to study the BEC-BCS crossover \cite{KC06}. 
The spatial resolution of this method is set by the size of the laser spot.
This cannot be made much less than  about 100$\mu$m, 
if the dipole forces arising 
from the gradients of the laser field are not to lead to significant trap losses 
\cite{Prodan03}.

\section{ionic spectrometer}

Recently, Kollath \textit{et.al.} \cite{Kollath07} 
proposed probing the local density of
cold atomic gases trapped in an optical lattice
using the coherent two-photon Raman photo-association of a single ion embedded in a cold gas with an atom 
of the gas. This could be used to measure the local density distribution of 
the atoms at different sites of an optical lattice as well as the spatial density 
correlation or single-particle correlation function in real time with a spatial resolution on  a 
nanometer scale.  Resolution of the hyperfine 
structure would also be possible. 

Here we demonstrate that the experiment proposed by Kollath \textit{et.al.} \cite{Kollath07} could be adapted 
to measure not only the density distribution (as in the original proposal by Kollath \emph{et.al}), but also 
the \emph{local single-particle energy distribution function} of a trapped Fermi gas \emph{in situ}---a 
long sought goal of the cold atom community. 
We show that, in a two-photon Raman spectroscopy measurement,
the photo-induced association rate of an ion held 
in a Paul trap in an atomic gas directly measures 
the local single-particle energy distribution of the atoms.
The spatial resolution is set by the  ``amplitude'' of the ion
oscillations in the harmonic Paul trap, which can be as small as 
10nm \cite{Wineland98}. 
 For example, it should be possible to
resolve a narrow region of possible FFLO phase in a Fermi gas with imbalanced
spin populations, and the variation of the density around a vortex
in a rotating Fermi gas. 
The spectral resolution is determined by the detunings 
and the Rabi frequencies of the transitions. For
Fermi gases prepared in an optical dipole trap near a Feshbach resonance, where
the s-wave interaction between fermions with different spins can
be tuned through the Feshbach resonance using a magnetic-field, we show 
that the photo-association time would resolve the variations in the 
single-particle energy distribution of the gas through the BEC-BCS crossover.



\begin{figure} [ht]
\centering
\includegraphics{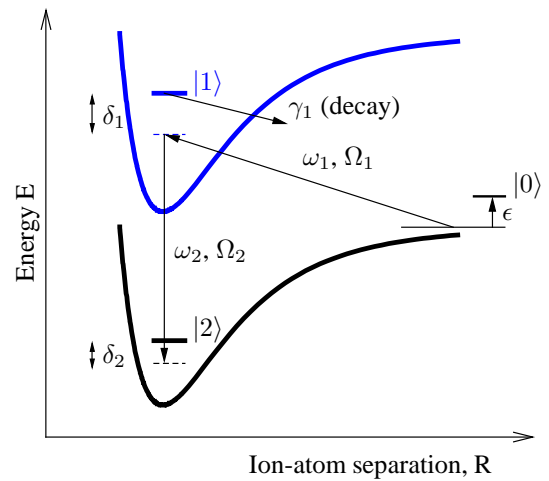}
\caption{\label{fig:Raman_coupling} 
The two-photon Raman coupling of an ion 
and an atom with energy $E$. 
Two lasers with frequencies $\omega_1$ and $\omega_2$ are incident 
on an atomic Fermi gas within which an ion is held in a Paul trap.
The curves show the interaction potential and the energy levels 
for the ion interacting with an atom as a function of the atom-ion
separation, $R$.
$|0\rangle$ is
the combined state of the ion and an atom 
(with energy $E$), $|1\rangle$ is an excited molecular state
and $|2\rangle$ is the ground state of the molecular ion.
Absorption of a photon with frequency $\omega_1$ 
leads to a virtual transition to the excited state,
which has a lifetime $1/\gamma_1$. The Rabi 
frequency for this coupling is $\Omega_1$.
Stimulated emission
at the frequency of the second laser leads to de-excitation into
the ground state of the molecule. The detuning from 
the respective resonances
(for an initial state with $E=0$) are $\delta_1$ and $\delta_2$.}

\end{figure}

The two-photon association process is illustrated schematically
in Fig \ref{fig:Raman_coupling}. 
A probe laser would  be used to sense the state of the ion
and the time for association recorded. Once association has been
detected, one of the lasers controlling the process is switched off to
prevent Rabi oscillations and the molecule decays back to a free
atom and ion. The association and decay processes are fast and 
the process can be repeated so that the accumulation of accurate statistics
should be straightforward. The time-limiting
step is likely to be the scattering rate within the Fermi gas as this
controls the rate at which any excess energy released in the decay process
can be dissipated. (The system needs to return to equilibrium before the next
measurement can be made.) Estimates based
on previous experimental studies \cite{Loftus02} suggest that this could be as short as a
few milliseconds if s-wave scattering is not forbidden, {\it i.e.}  
there are unpolarized spin populations.

We consider an ion located in a Paul trap surrounded by a degenerate atomic
Fermi gas. To a good approximation the ion is affected only by the Paul 
trap, and the atoms are confined only by the optical trap allowing for
effective independent control of the two components \cite {Kollath07}.
Rapid scanning of the ion trap is known to be possible with displacements
of 1.2mm within $50\mu$s without exciting vibrational modes \cite {Rowe02}.  
The system is illuminated by two lasers of frequencies $\omega_{1}$ and
$\omega_{2}$ and intensities $I_{1}$ and $I_{2}$. The energy level
scheme is shown in Fig. \ref{fig:Raman_coupling}. The atom-ion system, initially in state
$|0\rangle$,  can absorb a photon with frequency  $\omega_{1}$ and,
if the Franck-Condon condition is satisfied, 
lead to the formation of a molecule in the excited state $|1\rangle$. 
The second laser induces a transition of the molecule to its ground state
$|2\rangle$ via stimulated emission. 

The association rate
is controlled by the laser frequencies and Rabi frequencies, and is 
a sensitive function of the energy of the atom $E$.
If the system were illuminated
continuously, subsequent dissociation of the molecule back to the ion and
free atom (state $|0\rangle$) could occur by absorption of a
photon  with frequency $\omega_{2}$. Stimulated emission
at frequency $\omega_1$ and damped Rabi
oscillations would then be observed. However these would make identification of
a well-defined initial state difficult. Instead
the laser with frequency $\omega_1$ should be
switched off as soon as association has been observed. The molecule 
will then decay spontaneously at a rate controlled by the inverse lifetime
of the state $|1\rangle$, $\gamma_{1}$. 
This dissociation process will leave the ion in a high vibrational state 
and the atom with a high enough energy to escape the atom trap. After
waiting for the system to re-equilibrate the process can be repeated.

To calculate the association rate we use the Bohn-Julienne  recipe \cite{Bohn96, Bohn99}. 
If the detuning $\delta_1 \gg \delta_2$ and $\delta_1 \gg \Omega_2$, where
$\Omega_2$ is the Rabi frequency for the oscillation $|2\rangle \leftrightarrow |1\rangle$,
the molecular level $|1\rangle$ is adiabatically eliminated \cite{Orszag}, 
and we can work with an effective two-level system. 
The interaction of the system with the lasers results in an optical Stark shift, $\eta$,
a broadening of the ground state molecular level, 
$\gamma_2$ \cite{Bohn96, Bohn99}, and a Rabi frequency for the whole process, $\Omega$, given by: 
\begin{equation}
\eta=\frac{\Omega_{2}^{2}}{\delta_{1}}, \hspace{0.3cm}
\gamma_{2}=\frac{\Omega_{2}^2}{\delta_{1}^2}\gamma_{1}, \hspace{0.3cm}
\Omega=\frac{\Omega_{1}\Omega_{2}}{\delta_1}.
\label{eq:two-level}
\end{equation}
The shift of the level $|0\rangle$ is
negligible provided $\Omega_{1} \ll \Omega_{2}$, where $\Omega_1$ is the Rabi frequency 
for the transition 
$|0\rangle \leftrightarrow |1\rangle$.
First, we consider a spin-polarized Fermi gas implying no interaction between the atoms.
The Hamiltonian for the effective two-level system is then
\begin{eqnarray}
H & = &  (\omega_{0}+\delta_{2}-\eta)b^{\dagger}b+
\sum_{\mathbf{p}}\epsilon a_{\mathbf{p}}^{\dagger}a_{\mathbf{p}}+ \sum_{n}E_{n}I_{n}^{\dagger}I_{n}  \nonumber \\
& &
- \left( \sum_{\mathbf{p}}\Omega(\mathbf{p})b^{\dagger}a_{\mathbf{p}}I_{0}e^{-i\omega_0 t}+h.c.\right),
\label{eq:Heff}
\end{eqnarray}
where $b$ 
is annihilation operator for the ground
state of the molecular ion, $a_{\mathbf{p}}$ is the annihilation operator
for an atom in state with momentum $\mathbf{p}$, 
$I_n$ is the annihilation operator
for the ion in $n$-th eigenstate with energy $E_n$ of the ion trap, 
$\epsilon=p^2/2m$ is the kinetic energy of the atom, $\omega_{0}=\omega_{1}-\omega_{2}$.
The Rabi frequency $\Omega$ depends on the initial energy of the atom 
$\Omega\propto\epsilon^{1/4}$ \cite{Bohn96, Bohn99}. 

Typical values for the zero-point wavefunction spread of the ion
or molecule in the Paul trap, $z_{0}$, are of the order of 10nm \cite{Wineland98} and
hence much less than the wavelengths of the incident lasers or the atoms in the trap: 
As a result, any recoil effects
resulting from the absorption and photon emission will be negligible.
A simple calculation yields an expression for the photo-association rate of an atom and the ion 
\begin{eqnarray}
\Gamma(\mathbf{R}) & = &\frac{2^{3/2}}{3\pi}R_{TF}^{3}m^{3/2}\int\sqrt{\epsilon} 
G(E)n_{\sigma}(\epsilon,\mathbf{R}) d\epsilon ,
\label{eq:rate} \\
G(E)  & = & \frac{\gamma_{2}\Omega(\epsilon)^{2}}
                        {(\tilde{\delta} - E)^{2} 
                              +\frac{\gamma_{2}^{2}}{4}}, \hspace{0.4cm} \tilde{\delta} = \delta_{2}-\eta .
\label{eq:res}
\end{eqnarray}
Here $R_{TF}$ is the Thomas-Fermi radius of the gas in the trap with 
frequency $\omega=(\omega_{x}\omega_{y}\omega_{z})^{1/3}$, $\sigma$ is the spin index, 
$m$ is the mass of a fermion. $ n_{\sigma}(\epsilon,\mathbf{R})$  
is the momentum distribution of the atoms at a point with position
$\mathbf{R}$. For non-interacting fermions the energy of the atom is $E=\epsilon$.  

\begin{figure} [ht]
\centering
\includegraphics[totalheight=0.22\textheight]{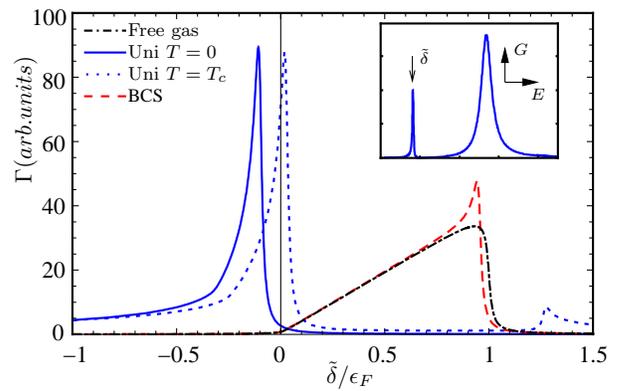}

\caption{\label{fig:G_functions}
The photo-association rate, $\Gamma$,  for an ion held in a Fermi gas 
as a function of the detuning parameter, $\tilde{\delta}$, computed from (\ref{eq:rate}) 
and
(\ref{eq:res}) 
for four different regimes
of the uniform Fermi gas: 
free fermions at $T=0$ (dashed-dotted line); the BCS regime with $k_F^0a=-0.5$ ($\Delta=0.047\epsilon_F$) at $T=0$,
where $k_{F}^{0}$ is the 
Fermi wave number and $a$ is the scattering length (dashed line); the unitarity limit (solid line) at $T=0$. We also present the photoassociation rate for the unitarity regime at $T=T_c$ (dotted line). 
The major peak is due to breaking of fermion pairs. The minor peak shifted to positive 
$\tilde{\delta}$ is attributed to thermally excited atoms. Inset: The photo-association rate density, $G(E)$, 
for the transition $|0\rangle \rightarrow |2\rangle$. 
$G(E)$ is shown for the three-level system 
assuming $\delta_1 = 2.5 \delta_2$ (see Fig \ref{fig:Raman_coupling}).
The separation of the two peaks $\xi=\sqrt{(\delta_1-\delta_2)^2 + 4 \Omega_2^2}$.}

\end{figure}

For the three level system of Fig \ref{fig:Raman_coupling}, 
the photo-association rate $G$ exhibits two peaks. 
This is shown in the inset of Fig. \ref{fig:G_functions}. The width of the peak at low energy is determined by the width
of the ground state level 
$\gamma_{2}$.  If the detuning $\delta_1 \gg \delta_2$ and $\delta_1 \gg \Omega_2$,
the peak at low energies can be made
very sharp (see (\ref{eq:two-level})) and the photo-association rate $\Gamma$ is determined by the local 
density of atoms with this energy.
The maximum of the function $G(E)$, given by (\ref{eq:res}), for the case
$\Omega_1 \ll \Omega_2$
is at $E = \tilde{\delta}$.
We find
\begin{eqnarray}
G(\tilde{\delta})=\frac{4\Omega^{2}}{\gamma_{2}} .
\end{eqnarray}


The presence of a
molecule would be detected by measuring the change of oscillation
frequency of the heavier molecular ion in the trap, or by observing
the absence of the resonance light scattered by the molecule. The available 
state-sensitive optical detection time is of the order 0.1ms \cite{Itano93,Sugiyama97}, so photo-association 
rates of the order  $10^{3}s^{-1}$ or less are achievable.
Once association is observed, switching off the laser with frequency $\omega_1$ 
will allow the molecular state $|2\rangle$ to decay spontaneously. 
As a result of the decay process, the atom
escapes from the optical dipole trap while the ion 
remains confined by the Paul trap. 
The ion may now
be in an excited vibrational state \cite{note_on_atomdecay}. However,
the rate of ion-atom collisions is estimated to be $\sim 10^{3}s^{-1}$ 
for the system under consideration \cite{Cote00}, which corresponds to a time-scale
significantly longer than the time on which the ion can be cooled ($\sim 200\mu$s) 
\cite{Monroe95,Wineland98}. 
The vibrational excitations of the ion should not therefore lead to 
much heating of the atoms of the system through collisions with the ion.  
Any energy, that is transferred to the atoms, would be redistributed 
among the atoms on a time-scale set by the energy-relaxation time in
the atomic gas.
In strongly interacting Fermi gases this has been observed to be around $1$ms \cite{Loftus02}
and even shorter in some systems \cite{Salomon08}. This relaxation time
and the photo-association time itself are the longest time-scales  
in the process we are proposing. As both should be comparable or less than 1ms,
a cycle time of a few milliseconds should
be easily achievable.

If the ion were embedded in a Bose gas, spontaneous capture of an atom would be 
significant. The rate of spontaneous capture for a realistic Bose system such as that 
described in  \cite{Cote02} is of order $\Gamma_{sp}=600 s^{-1}$, which is comparable with 
the photo-association rate. This would make the use of this method to study
Bose gases problematic. For Fermi gases the spontaneous capture is suppressed by Fermi statistics 
and can be neglected.

Fitting the measured photo-association rate to the theoretical form
could be used, for example, 
to determine the gap of the gas as a function of the 
position 
to monitor the state of the system as it is tuned through
the BCS-BEC crossover.
In Fig. \ref{fig:G_functions}, we show the photo-association rate, 
obtained from (\ref{eq:rate}), for 
a Fermi system in the BCS and unitarity regimes for uniform gas at zero temperature.
We have used the mean field result for the zero-temperature distribution function 
for a system of interacting particles with equal spin 
populations in the local density approximation \cite{Viverit04}:
$
n(\epsilon, \bf {R})=\frac{1}{2}\left(1-\frac{\epsilon-\mu(\bf{R})}{\sqrt{(\epsilon-\mu(\bf{R}))^2 
+\Delta(\bf{R})^2}}\right) .
$
Here $\Delta(\bf{R})$ and $\mu(\bf{R})$
are the local values of the order parameter and chemical potential. The energy of interacting atoms is $E=\mu-\sqrt{(\epsilon-\mu)^2+\Delta^2}$.
Generalizations to the case of non-zero temperatures \cite{Perali04,He05} 
and to the imbalanced population case 
\cite{Combescot01,Chen08} are straightforward.
$\mu$ and $\Delta$  are determined by the scattering length $a$
via the usual gap equation, which at $T=0$ is:  
$
\frac{m}{4\pi a}= \int \frac{d{\bf p}}{(2\pi)^3}\left( \frac{m}{p^2} - 
\frac{1}{2\sqrt{(\epsilon-\mu)^2+\Delta^2}}\right) 
$ 
and the density equation. 

We also plot in Fig. \ref{fig:G_functions} the photoassociation rate for the unitarity regime at $T=T_c$. To calculate the association rate we used the strict mean field theory, where finite-time effects were neglected \cite{Chen08}. The main peak is associated with the breaking of fermion pairs.  The second peak shifted to the positive detunings is due to thermally excited fermions.
The differences between the curves shown in Fig. \ref{fig:G_functions} 
should be visible in experiment thereby allowing direct identification of the
state of the Fermi gas both as a function of the interaction parameters and
of the position in the trap.

\section{Discussion}
Now we compare the method proposed in this paper with the RF spectroscopy \cite{He05}. In the proposed scheme the momentum of the atom before and after association is not conserved because the Fermi energy of the atoms $\epsilon_F\sim 20kHz$ is much less than the frequency of the Paul trap $\omega=7.8MHz$, which is required to trap a single ion \cite{Wineland98}. This leaves the molecular ion in the same vibrational state after association, with the momentum of the atom absorbed by the trap. On the other hand the momentum of the atom is conserved in the RF spectroscopy, because the atom experiences the transition to a free state after excitation. Thus, the detuning corresponding to the maximum of the RF current in RF spectroscopy is always negative, while the detuning corresponding to the maximum of the photoassociation rate in our proposal can be either negative or positive (see Fig. \ref{fig:G_functions}). We also comment on the final-state interaction. In RF spectroscopy an atom excited by an RF pulse interacts with the ground-state atoms via a Feshbach resonance, which leads to corrections to the RF spectrum. In our case this effect can be neglected because of the dispersion interaction between the ion (molecule) and the atoms, which is much stronger than the final-state interaction. 


Finally, we turn our attention to the many body interaction between the ion and the gas. 
When the ion and an atom form a bound state, one might expect to observe the manifestations of the many-body 
interaction such as the Fermi-edge singularity (FES) \cite{ND69}. 
The photo-association rate for a uniform system of non-interacting
fermions at zero temperature should be
\begin{eqnarray}
\Gamma & \propto & {\Theta(\epsilon_{F}-\tilde{\delta})(\epsilon_{F}-\tilde{\delta})^{\alpha-1}}
\label{eq:fes}
\end{eqnarray}
with the exponent $\alpha$ determined by the difference in the phase shift (modulo $\pi$) 
at $\epsilon_{F}$ for fermions scattering off a free ion and a molecular ion. 
The singularity arises because of the multiple low energy excitations 
near the Fermi surface as a result of the sudden switching of the potential as seen,
for example, when a core-hole state is created as the result of an x-ray absorption process in a metal.
In our case the change of the potential, as a result of the 
formation of bound state, is not significant. This is because the atom does not screen the 
potential of the ion.  As a  result  $\alpha\approx 1$ in (\ref{eq:fes}), and (\ref{eq:rate}) 
gives the correct association rate. 

\section{conclusions}
We have proposed an ionic-spectrometer  to measure the \textit{local single-particle energy distribution} 
of the degenerate Fermi gases with an energy resolution determined by the laser detunings and a spatial 
resolution on the nanometer scale. We discuss \textit{in situ} experiment to measure the pairing gap of an ultracold Fermi gas locally.

\begin{acknowledgements}
The authors thank Michael K\"ohl for many helpful discussions.
The work was supported by EPSRC-EP/D065135/1. 
\end{acknowledgements}

\bibliographystyle{apsrev}
\bibliography{fermometer2}

\begin{thebibliography}{42}
\bibitem[{not()}]{note_on_atomdecay}
\bibinfo{note}{Measuring the loss rate of molecules due to the spontaneous
  decay in a Fermi gas has been suggested as a spectroscopy \cite{KC06}. There,
  the atoms' high kinetic energy after dissociation means that they both escape
  the trap cleanly. Here it is easier to concentrate on the association
  process, because of the vibrational state of the ion after dissociation.}
\end{thebibliography}

\end{document}